\documentclass[aps,prb,twocolumn,showpacs,groupedaddress]{revtex4}

\usepackage{color}
\usepackage{graphicx}

\begin{document}

\title{Spin-Dynamics of the antiferromagnetic $S$=1/2-Chain at finite magnetic Fields and intermediate Temperatures}

\author{S. Grossjohann and W. Brenig}
\affiliation{Institut f\"{u}r Theoretische Physik, TU Braunschweig, 38106 Braunschweig, Germany\\
}

\date{\today}

\begin{abstract}
We present a study of the dynamic structure factor of the antiferromagnetic spin-1/2 Heisenberg chain at finite temperatures and finite magnetic fields. Using Quantum-Monte-Carlo based on the stochastic series expansion and Maximum-Entropy methods we evaluate the longitudinal and the transverse dynamic structure factor from vanishing magnetic fields up to and above the threshold $B_c$ for ferromagnetic saturation, as well as for high and for intermediate temperatures. We study the field-induced redistribution of spectral weight contrasting longitudinal versus transverse excitations.
At finite fields below saturation incommensurate low-energy modes
are found consistent with zero temperature Bethe-Ansatz.
The crossover between the field induced ferromagnet above $B_c$ and the Luttinger liquid below $B_c$ is analyzed in terms of the transverse spin-dynamics. Evaluating sum-rules we assess the quality of the analytic continuation and demonstrate excellent consistency of the Maximum-Entropy results.
\end{abstract}


\maketitle

\section{Introduction}
The antiferromagnetic spin-$1/2$ Heisenberg chain (AFHC) is one of the most
intensively studied strongly correlated quantum many body systems. In the
presence of an external magnetic field, its generalization to anisotropic
exchange, the XXZ model, reads
\begin{equation}\label{eqn1}
H=J\sum_l [
S^z_lS^z_{l+1} +
\frac{\Delta}{2}(S^+_lS^-_{l+1}+S^-_lS^+_{l+1})
- B S^z_l ],
\end{equation}
where $J$ is the antiferromagnetic exchange interaction with an anisotropy
ratio $\Delta$, $S^{\pm,z}_l$ are the spin operators on site $l$ of the
chain, and $B=g\mu_B\hbar H$ is the magnetic field. From a materials
perspective SrCuO$_2$ \cite{Ishida1994a} ($J/k_B\approx 2600K$),
Sr$2$CuO$_3$ \cite{Motoyama1996a,Ami1995a,Takigawa1996a} ($J/k_B\approx
2200K$), and Cu\-(C$_4$\-H$_4$\-N$_2$)\-(N\-O$_3$)$_2$ \cite{Losee1973a}
($J/k_B\approx 10.7K$) are topical examples of both, low- and high-$J$ AFHC
compounds which have been studied intensively \cite{Broholm2002a}.
Recently, dynamical correlation functions of the AFHC have become
accessible to a variety of high resolution spectroscopies at finite
temperature and in the presence of external magnetic fields, eg. inelastic
neutron scattering (INS) \cite{Stone2003a}, high-field nuclear magnetic
resonance (NMR) \cite{Takigawa1997a,Thurber2001a,Wolter2005a,Kuhne2008a},
muon spin-resonance ($\mu$SR) \cite{Pratt2006a}, and magnetic transport
\cite{HeidrichMeisner2007a,Sologubenko2007a}.

The AFHC is integrable, including the cases of $h\neq 0$ and $\Delta\neq
1$. Bethe-Ansatz (BA) \cite{Bethe1931a,Faddeev1981a} has been used to
investigated its ground state properties. Static thermodynamic quantities,
eg. the specific heat, the magnetic susceptibility and the magnetization
have been investigated by several methods including thermodynamic BA,
Quantum Monte Carlo (QMC), as well as transfer-matrix and density-matrix
renormalization (DMRG) group, see
\cite{Klumper1993a,Johnston2000a,Klumper2000a} and refs. therein.

Fractionalization of the spin excitations into multi spinon states is a
fingerprint of the AFHC \cite{CloizauxPearson62,Yamada1969a}. At zero
temperature, $T=0$, numerical analysis of these excitations has been
carried out in many studies using exact diagonalization (ED) of finite AFHCs,
see eg. refs. \cite{Muller1981a,Bonner1982a,Lefmann1996}, including the
effects of $h$ and $\Delta$, as well as by dynamical variants the DMRG
\cite{Kuhner1999,Gobert2005a}. In principle, also BA allows
to determine dynamical correlation functions, however calculating the corresponding matrix elements is highly non-trivial and progress has been made only recently. By now analytic expressions for
dynamical spin correlation functions are available for the two-
\cite{Bougourzi1996a,Karbach1997a,Bougourzi1998a} and the four-spinon
sector \cite{Bougourzi1996b,Abadaa1997a,Lakhala2005a,Caux2006a} at
$\Delta=1$, $h=0$ and $T=0$. In addition, determinant approaches
\cite{Kitanine2000,Caux2005a} allow for numerical treatment of two-
\cite{Biegel2002a,Biegel2003a,Sato2004a} and many-spinon
\cite{Karbach2000a,Caux2005b} states of the XXZ chain in finite magnetic
fields, at $T=0$. Finally, mapping to field theory in the continuum limit
\cite{Schulz1986a} has been used to study the small-$q$ behavior of
longitudinal dynamical structure factor in the gapless regime
\cite{Pustilnik2006,Pereira2006a,Pereira2007a}.

At finite temperatures, the dynamical correlations functions of the AFHC
remain an open issue. The dynamical structure factor
$S^{\alpha\beta}(q,\omega)$ has been studied by complete ED of small systems \cite{Fabricius1997a,Fabricius1998a} in
the context of spin diffusion, see
\cite{HeidrichMeisner2003a,HeidrichMeisner2007a}. and refs. therein.
However, such analysis is limited by finite size effects to $k_B T\gtrsim
J$. Recently, finite temperature real-time auto- and next-nearest neighbor
correlation functions have been accessed by DMRG methods
\cite{Sirker2005a,Sirker2006a}. However, the time range of such
calculations is limited, as the spectrum of the reduced density matrix used
to truncate the Hilbert space becomes dense. In this respect QMC remains a
key tool to evaluate the $S^{\alpha\beta}(q,\omega)$, for system sizes
which are close to the thermodynamic limit, over the complete Brillouin
zone, and at finite temperatures, with the limitations set primarily by the
analytic continuation of imaginary-time data \cite{Jarrell1996}. QMC
analysis of $S^{\alpha\beta}(q,\omega)$ has been carried out for $h=0$
\cite{Deisz1990a,Deisz1993a,Starykh1997a}, results for $h\neq 0$, however,
are lacking.

The purpose of this work is to shed more light the
finite-temperature dynamical structure factor of the AFHC in the presence
of external magnetic fields using QMC. The paper is organized as follows.
In section \ref{B} we briefly summarize the QMC approach we use. Section
\ref{C} we analyze the transverse and longitudinal structure factor
versus temperature and magnetic field. In section \ref{D} we consider
several sum rules of the AHFC. We summarize and conclude in section
\ref{E}.

\section{Methods}
\label{B}
In this paper we present results for the Fourier transform of the dynamic structure factor
\begin{equation}
S^{\alpha\beta}(q,\omega) =
\int_{-\infty}^{\infty} dt \, e^{i\omega t} S^{\alpha\beta}(q,t)\,\, ,
\end{equation}
where $t$ refers to real time and $S^{\alpha\beta}(q,t)=\langle S^\alpha_q(t)S^\beta_{-q}\rangle$ with $S^\alpha_q=\sum_r e^{-i q r} S^\alpha_r$ being the spin component $\alpha$ at momentum $q$. In the following we discuss both, the {\it longitudinal} and the {\it transverse} structure factor, i.e. $\alpha\beta=zz$ and $\alpha\beta=xx$. These components are directly accessible to unpolarized neutron scattering. Other types of transverse components, eg. $\alpha\beta=+-$ require polarized neutrons and will not be considered here \cite{Grossjohann-unpublished}. For $\alpha\beta=xx(zz)$  $S^{\alpha\beta}(q,\omega)$ is related to the imaginary time structure factor $S^{\alpha\beta}(q,\tau) = \langle S^\alpha_q(\tau)S^\beta_{-q}\rangle$ through the integral transform
\begin{equation}
S^{\alpha\beta}(q,\tau)=\frac{1}{\pi}\int_{0}^{\infty}d\omega K(\omega, \tau)S^{\alpha\beta}(q, \omega)
\label{eqn:continuation}
\end{equation} 
with the kernel $K(\omega, \tau)=e^{-\tau\omega} + e^{-(\beta-\tau)\omega}$.

The imaginary time correlation functions $S^{\alpha\beta}(q,\tau)$ will be evaluated by QMC. Here we employ the stochastic series expansion (SSE) introduced by Sandvik {\it et al.} \cite{sandvik91}. This method is based on a particular form of the high temperature series expansion of the partition function which can be sampled efficiently \cite{sandvik02}. To this end, the Heisenberg model is rewritten in terms of $N_b$ bond operators
\begin{equation}
H=-J\sum_{b=0}^{N_b} (H_{1,b} + H_{2,b} )
\end{equation}
where $H_{1,b} = 1/2 - S^{z}_bS^{z}_{b+1}$ and $H_{2,b} = ( S^{+}_bS^{-}_{b+1} + S^{-}_bS^{+}_{b+1})/2$ refer to the {\it diagonal} and {\it off-diagonal} parts of $H$ within the $S^z$ basis. Within this notation
the partition function is
\begin{equation}
Z=\sum_{\alpha}\sum_{n}\sum_{S_n}\frac{(-\beta)^n}{n!}\left\langle \alpha \right| \prod_{k=1}^n H_{a_k, b_k}\left|\alpha\right\rangle \, \label{eq:partition}
\end{equation} 
where $|\alpha\rangle = \left|S^z_1,\ldots, S^z_N\right\rangle$ refers to the $S^z$ basis and $S_n$
\begin{equation}
S_n = [a_1, b_1][a_2, b_2]\ldots[a_n, b_n]
\end{equation}
is an index for the {\it operator string} $\prod_{k=1}^n H_{a_k, b_k}$,
labeling each specific product of operators where
$a_k\in \left\lbrace 1,2\right\rbrace$ and  $b_k\in\left\lbrace 1, \ldots N_b \right\rbrace$.

The operator string is subject to importance sampling by a Metropolis scheme, splitted into two different types of updates: a {\it diagonal} update which changes the number of diagonal operators $H_{1, b_k}$ in the operator string and a cluster type update which performs changes of the type $H_{1, b_k} \leftrightarrow H_{2, b_k}$. On bipartite lattices the latter, so-called {\it loop} update guarantees an even number of off-diagonal operators $H_{2, b_k}$ in the expansion. This ensures positiveness of the transition probabilities.

Imaginary time spin-correlation functions $S^{\alpha\beta}_{i,j}(\tau)= \langle e^{\tau H}S^{\alpha}_i e^{-\tau H}S^{\beta}_j\rangle$, for $\tau \in [0, \beta )$, and lattice sites $i,j$ can also be sampled by the SSE using that \cite{sandvik92}
\begin{equation}
S^{\alpha\beta}_{i, j}(\tau)= \left\langle \sum_{m=0}^{n}\frac{\tau^m(\beta-\tau)^{n-m}n!}{\beta^n(n-m)!m!}\overline{C}^{\alpha\beta}_{i, j}(m) \right\rangle_W \,
\label{w1}
\end{equation}
where, in contrast to the thermal average $\langle\ldots\rangle$, the brackets $\langle\ldots\rangle_W$ with subscript $W$ refer to averaging over operator string configurations with weights generated by the Metropolis scheme. The quantity $\overline{C}^{\alpha\beta}_{i, j}(m)$ in eqn. (\ref{w1}) is the static
real space correlation function
\begin{equation}
\overline{C}^{\alpha\beta}_{i, j}\left( m \right) = \frac{1}{n+1}\sum_{p=0}^{n}S^{\alpha}_{i}(p)S^{\beta}_{j}(p+m) \,
\label{eqn:corr}
\end{equation}
which, in the case $\alpha\beta = zz$, can be measured within the diagonal or slightly more efficient within the loop update \cite{troyerdorneich} where $m,p\leq n$ refer to positions within the operator string and $S^{\alpha}_{i}(p)$ refers at the intermediate state $|\alpha(p)\rangle =\prod_{k=1}^pH_{a_k, b_k}|\alpha\rangle$ of the expansion. Transverse correlations $\alpha\beta = xx, +-, -+$ in a code working in the $S^z$ basis can only be accessed by following the loop within the extended configuration space (see \cite{troyerdorneich}). Finally, the Fourier transform
$S^{\alpha\beta}(q,\tau)=\sum_{j}e^{-iq j}S^{\alpha\beta}_{j,0}(\tau)/N$ is used as input for the inversion problem eqn. (\ref{eqn:continuation}).

Extracting $S^{\alpha\beta}(q, \omega)$ from eqn. (\ref{eqn:continuation}) is notoriously complicated because of the QMC noise and the singular nature of $K(\omega, \tau)$. However, this problem can be handled successfully by Maximum Entropy methods. Here we use Bryan's algorithm \cite{bryan} which is specifically designed for over-sampled data sets and therefore well-suited to treat QMC results. For details we refer the reader to the Appendix.

\begin{figure*}[!t]
\begin{center}
\includegraphics{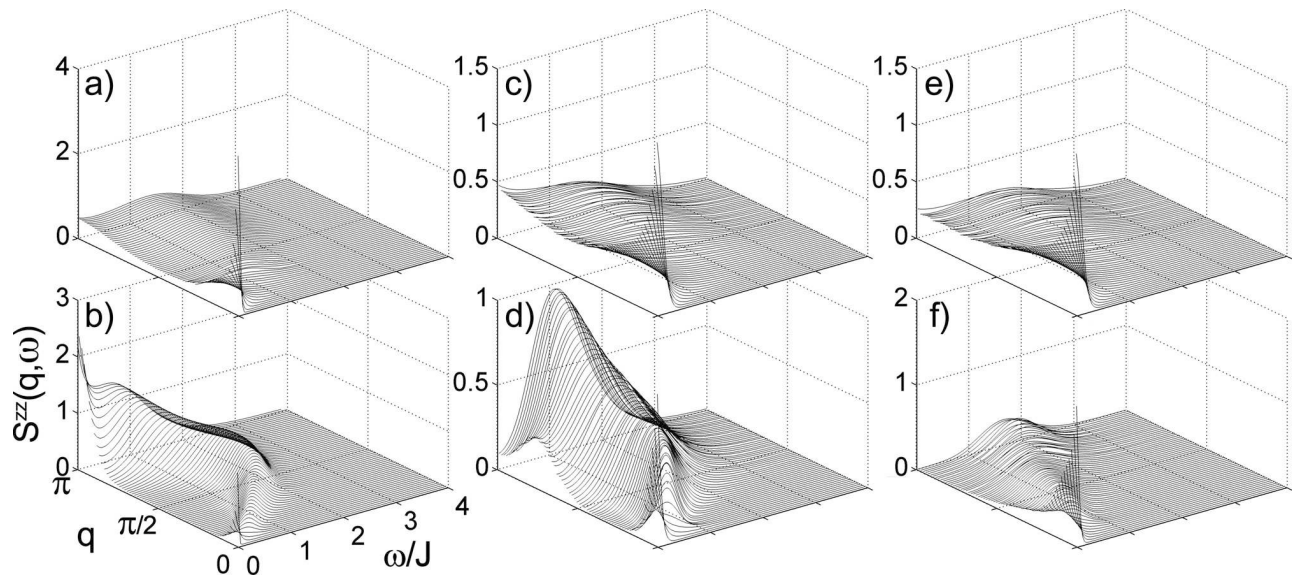}
\end{center}
\caption[1]{3D plot of the longitudinal dynamic structure factor by QMC + MaxEnt as function of frequency $\omega$ and wave vector $q$. Temperatures and magnetic field in units J: a) T=1 and B=0, b) T=0.25 and B=0, c) T=1 and B=1, d) T=0.25 and B=1, e) T=1 and B=2 and f) T=0.25 and B=2.}
\label{longitudinal_3d}
\end{figure*}
\begin{figure}[!t]
\begin{center}
\includegraphics{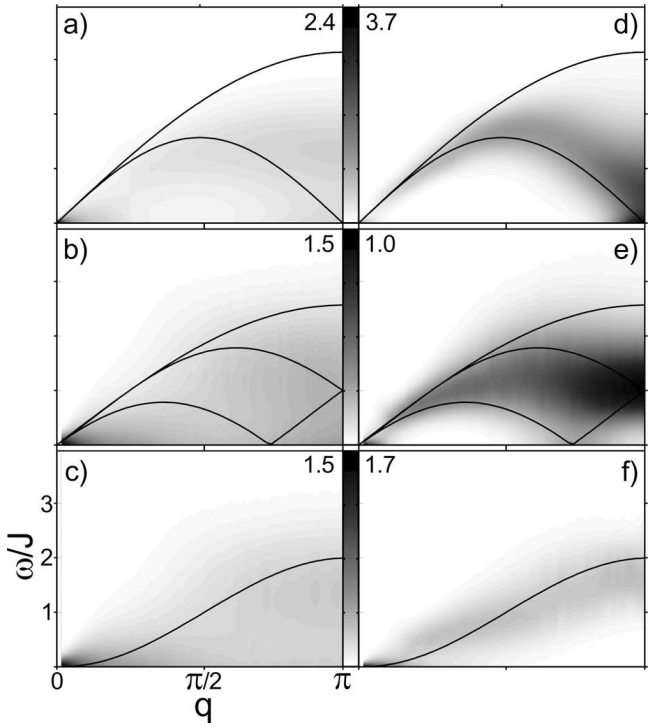}
\end{center}
\caption[1]{Contour plot of the longitudinal dynamic structure factor as function of frequency $\omega$ and wave vector $q$. Temperatures and magnetic fields in units of J: a) T=1 and B=0, b) T=0.25 and B=0, c) T=1 and B=1, d) T=0.25 and B=1, e) T=1 and B=2 and f) T=0.25 and B=2. For $B<B_c$ the solid lines are zero temperature excitation boundaries by the M\"uller-ansatz \cite{Muller1981a} while at critical fields the exact zero temperature $1-\cos(q)$ dispersion \cite{groen_longitudinaldispersion} is shown.}
\label{longitudinal_contour}
\end{figure}
\begin{figure}[!t]
\begin{center}
\includegraphics{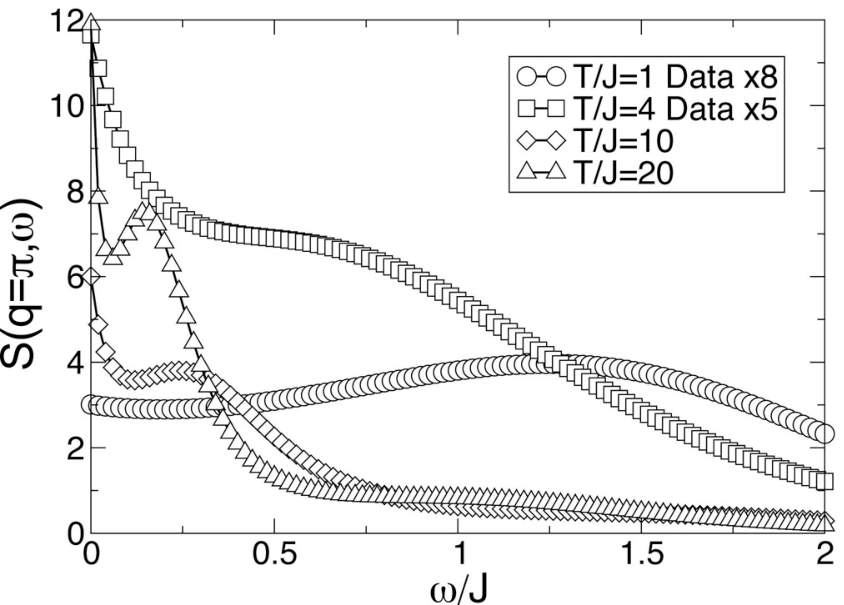}
\end{center}
\caption[1]{Zero-field dynamic structure factor at $q=\pi$ for four different temperatures (in units of J) $\left\lbrace 1, 0.25, 0.1, 0.05 \right\rbrace$. As the temperature decreases we find an increased divergent behaviour for $\omega\rightarrow 0$ as predicted by 2-spinon calculations. In addition there is a low frequency peak which shifts to lower energies while steadily gaining sharpness. Note that the data-set for $T=1$ and for $T=0.25$ was multiplied by a factor of eight, respectively five for illustrative reasons.}
\label{k64}
\end{figure}

\section{Results}
\label{C}
In this chapter we will present results for the transverse and longitudinal
dynamic structure factor at finite temperatures, in the range of
$T=J/20\ldots J$ and magnetic fields below and above the saturation field
$B_c$. All QMC calculations refer to systems with 128 sites, typically with
one billion Monte-Carlo updates (one diagonal and sufficient \cite{sandvik02} loop updates), distributed over 1000 bins. Only 50-100
$\tau$-points were extracted for each temperature in order to prevent
over-sampling of the relatively short expansion orders at elevated
temperatures close to $T=J$. An indication for over-sampling
is given by diagonalizing the covariance matrix which exhibits vanishing
eigenvalues in case of statistically dependent data.

\subsection{Longitudinal Dynamic Structure Factor $S^{zz}(q,\omega)$}

In Fig. \ref{longitudinal_3d} and \ref{longitudinal_contour} we show the
longitudinal dynamic structure factor both, as a 3D and a contour plot for
two different temperatures $T=\left\lbrace J, J/4\right\rbrace$ and three
different magnetic fields $B = \left\lbrace 0, B_c/2, B_c\right\rbrace$. The
solid lines displayed in the contour plots for $B<B_c$ refer to the upper and lower
boundaries of the two-spinon spectrum as obtained from BA
selection rules \cite{Muller1981a}. For zero magnetic field they enclose a region which, within 2-spinon calculations, contains about 73\% of the zero temperature spectral weight \cite{Karbach1997a}.  We will now focus on each of the
magnetic fields separately.

\subsubsection{The case $B=0$}

At zero magnetic field and high temperatures, i.e. Figs.
\ref{longitudinal_3d}a) and \ref{longitudinal_contour}a), we find a strong
broadening of spectral features. While the region of finite spectral weight
remains bounded from above by $J\pi\sin|q/2|$ \cite{yamada69}, significant
weight appears below the lower two-spinon boundary $\frac{\pi
J}{2}\sin|q|$ set by de Cloizeaux-Pearson \cite{dCP}. Most
noteworthy, high spectral weight occurs for
$q,\omega\rightarrow 0$. This intensity is related to
spin conservation which dominates the long wave-length
dynamics in the quasi-classical regime $k_B T\gg J$.
The question whether the long wave-length spin dynamics in the AFHC can be described by spin {\it diffusion} is a long-standing issue with no final answer as of today. For a recent review on the present status and related referencs we refer to \cite{HeidrichMeisner2007a}.

Unfortunately QMC is too sensitive to the default model for the MaxEnt continuation in the small-$q,\omega$ regime \cite{Deisz1993a} to elucidate the issue of spin-diffusion. Yet, we would like to mention agreement of our results regarding the frequency-transformed autocorrelation function $S^{zz}_0(\omega)$, i.e. the q-integrated dynamic structure factor (not shown within this work) with previous QMC, performed at $B=0$, high temperature series expansion \cite{Starykh1997a} and TMRG \cite{Sirker2006a}. These results exhibit a $\omega^{-0.3\ldots -0.4}$-divergent behaviour which bears resemblance to the phenomenological approaches by Bloembergen \cite{bloembergen1949} and de Gennes \cite{gennes1958} who predicted $\omega^{-1/2}$.

Next we consider lower temperatures, i.e. $T=J/4$. As is obvious from Figs.
\ref{longitudinal_3d}b) and \ref{longitudinal_contour}b),
spectral weight is removed from the long wave-length regime in this case.
Both figures demonstrate that most of the spectral weight is confined within
the two-spinon boundaries with however still an appreciable intensity below
the lower boundary. This is consistent with findings reported in
\cite{Deisz1993a}. In contrast to $T=J$ we find a strongly enhanced spectral
weight at $q=\pi$ owing to the increase of the antiferromagnetic correlation length \cite{kimwiese} which is consistent with the autocorrelation function reported in ref. \cite{Sirker2006a}.

In the limit $(q,\omega)\rightarrow (\pi,0)$ we find indications for
diverging behaviour of $S^{zz}(\pi,\omega)$ with decreasing temperatures.
This is shown in more detail in Fig. \ref{k64}, scanning a wide range of
temperatures from $T=J$ to $T=J/20$. As can be seen, the spectrum consists
of an upturn for $\omega\rightarrow 0$ and a peak at finite $\omega$. The
latter peak shifts to lower energies while gaining sharpness as
$T\rightarrow 0$. For $T\rightarrow 0$, Fig. \ref{k64} suggest that the
peak will merge with the zero-$\omega$ upturn to form a single divergence
at $\omega\rightarrow 0$, as predicted by 2-spinon calculations at $T=0$
which lead to $S^{zz}(\pi,\omega)\sim\omega^{-1}$
\cite{Muller1981a}. A similar peak at finite $\omega$ was observed also in
ref. \cite{Deisz1993a}. However, smaller systems sizes in that case, i.e.
$N=32$, render the zero-$\omega$ upturn into a shoulder only. Biasing the
default model by several sum-rules, it was shown in ref. \cite{Deisz1993a},
that $S^{zz}(\pi,\omega)$ on 32-site systems could be obtained with only
a single peak at finite $\omega$. Recent SSE-QMC on 128-site systems at
$B=0$ show only a single rounded maximum, centered at $\omega=0$ \cite{Starykh1997a}. While all these findings are consistent with the formation of a zero-$\omega$ divergence as $T\rightarrow 0$, they show that the details of the low-$\omega$ spectrum are subject to details of the
MaxEnt approach. Nevertheless, we will detail later that our results are
consistent with several sum-rules, including those which are particularly
sensitive to the low frequency behavior of the spectrum.

\begin{figure*}[!t]
\begin{center}
\includegraphics{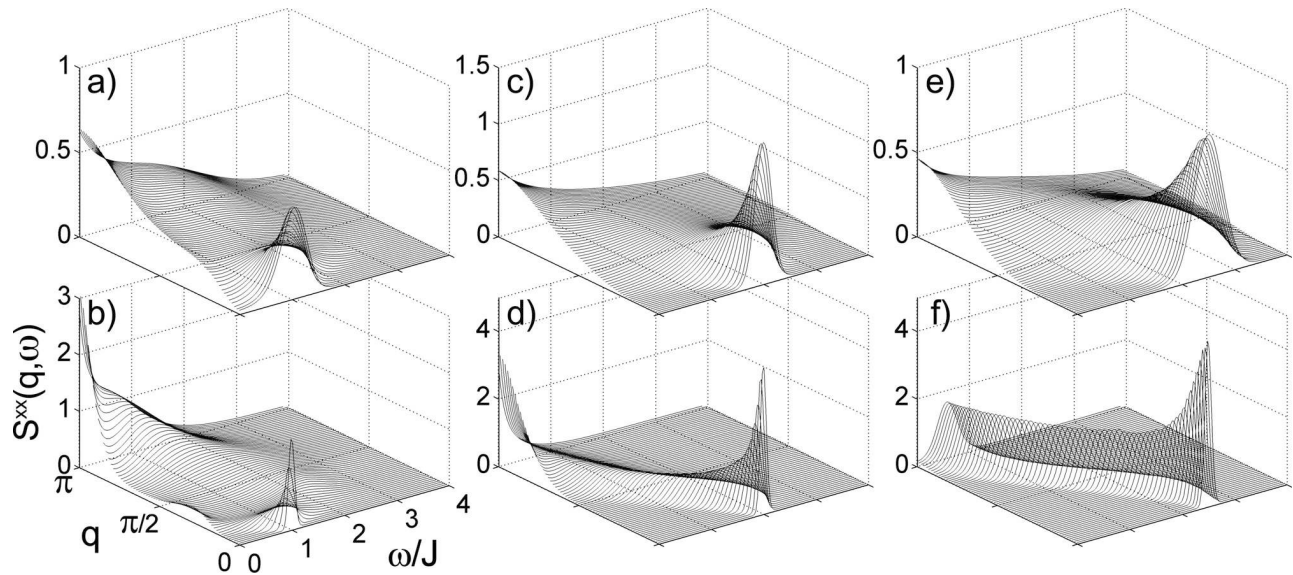}
\end{center}
\caption[1]{Transverse dynamic structure factor by QMC + MaxEnt as function of frequency $\omega$ and wave vector $q$. Temperature and magnetic field in units of J: a) T=1 and B=1, b) T=0.25 and B=1, c) T=1 and B=2, d) T=0.25 and B=2, e) T=1 and B=2.5 and f) T=0.25 and B=2.5.}
\label{transverse_3d}
\end{figure*}

\subsubsection{The case $B=B_c/2$}

Figs. \ref{longitudinal_3d},\ref{longitudinal_contour} c) and d) depict the
longitudinal structure factor at half of the critical field. The impact of a
finite magnetic field is fourfold. First, at zero momentum the
longitudinal
structure factor is proportional to the square of the field-induced
magnetization at zero frequency, i.e. $S^{zz} (q=0,\omega)\sim \langle
S^z\rangle^2 \delta(\omega)$. To focus
on the remaining spectrum, we have
chosen to skip the single wave vector $q=0$ in all 3D, as well as contour
plots of $S^{zz} (q,\omega)$ for $B\neq 0$. Second, longitudinal
excitations with $q\neq 0$ will have decreasing matrix elements with
increasing magnetic field. This is consistent with the evolution of the
overall scale in Fig. \ref{longitudinal_3d} a)-e) and b)-f).
Third, longitudinal spin-excitations at the zone
boundary are energetically unfavorable in a magnetic field.
In fact, at low temperatures a gap can be observed at $q=\pi$,
which is proportional to the
magnetic field \cite{Muller1981a} (see Fig. \ref{longitudinal_contour} d)).
Finally, a soft mode occurs at an
incommensurable wave vector $q_s = \pi(1-2\langle S^z\rangle)$ (see Fig.
\ref{longitudinal_contour} d)). This can be
understood in terms of the Jordan-Wigner fermionic description of the
AFHC \cite{jordanwigner1, jordanwigner2, jordanwigner3}, where $S^z_q$ is
related to the fermion density and the magnetic field plays the role of a
chemical potential driving incommensurability. This finding is consistent
with ref. \cite{Muller1981a}, with interacting spin-wave calculation
\cite{johnson1986} as well with finite system diagonalization
\cite{parkinson1985}. The role of temperature is evident. At high
temperatures, i.e. $T=J$ in Fig. \ref{longitudinal_contour} c),
$S^{zz}(q,\omega)$ is rather featureless and extends clearly beyond the
boundaries set by the two-spinon continuum. This changes as the temperature
is lowered to $T=J/4$, Fig. \ref{longitudinal_contour} d), where the
spectrum is far more confined to within the dCP boundaries and displays more
pronounced features. In particular, the weight is enhanced as
$(q,\omega)\rightarrow (\pi,J)$.

\subsubsection{The case $B=B_c$}

For $B\geq B_c$ and $T=0$, the statistical operator of the AFHC is pure and
corresponds to the fully polarized state, i.e. $S^{zz}(q,\omega)=N (1/4)\,
\delta_{q,0} \, \delta(\omega)$. Additional finite spectral weight for
$q,\omega\neq 0$ will occur only for $T>0$. To observe this we have again
removed the wave vector $q=0$ from Figs. \ref{longitudinal_3d},
\ref{longitudinal_contour} e), f), which are at the critical field. Indeed,
on lowering the temperature from panel e) to panel f) in Fig.
\ref{longitudinal_3d}, the remaining total spectral weight decreases. Apart from
this the higher temperature spectrum is rather featureless, while the lower
temperature spectrum clearly resembles the exact zero temperature dispersion
of $(1-\cos(q))$ \cite{groen_longitudinaldispersion} (see Fig. \ref{longitudinal_contour}
f)). This excitation has a constant spectral weight $2\pi/N$ for $q\neq0$, which vanishes in the thermodynamic limit.

\subsection{Transverse dynamic structure factor $S^{xx}(q,\omega)$}

In Fig. \ref{transverse_3d} and \ref{transverse_contour} we show the
transverse dynamic structure factor as 3D and as contour plots for identical
temperatures $T/J=\left\lbrace 1, 0.25\right\rbrace$ as for the longitudinal
dynamic structure factor, however for a different range of magnetic fields
$B/J = \left\lbrace 1, 2, 2.5 \right\rbrace$. For vanishing magnetic field
we refer to
Figs. \ref{longitudinal_3d}, \ref{longitudinal_contour} a), b) for
$S^{xx}(q,\omega)$ which is identical to $S^{zz}(q,\omega)$ at
$B=0$ due to $SU(2)$ invariance.

\subsubsection{The case $B=B_c/2$}

\begin{figure}[!t]
\begin{center}
\includegraphics{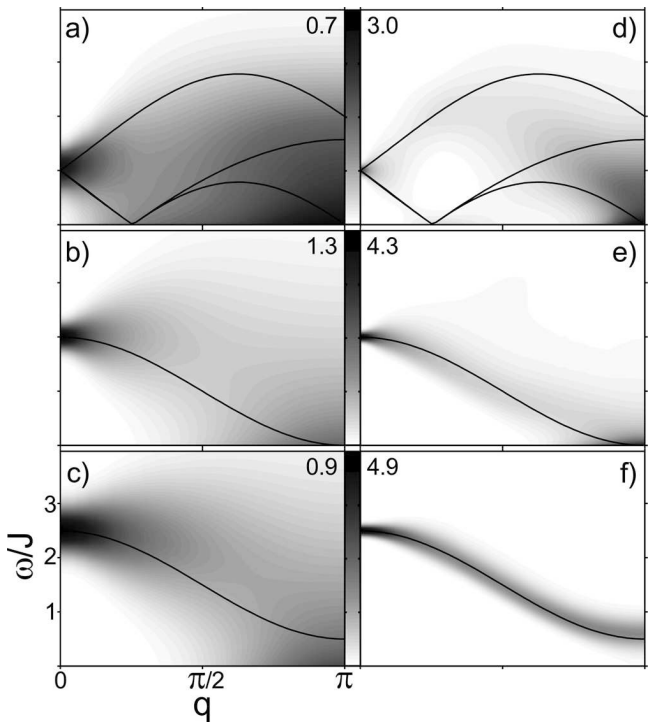}
\end{center}
\caption[1]{Transverse dynamic structure factor by QMC + MaxEnt as function of frequency $\omega$ and wave vector $q$ in a contour plot. For parameter details see text or Fig. \ref{transverse_3d}. The solid lines for half critical field a) and b) are zero temperature excitation boundaries of different BA selection rules (see \cite{Muller1981a}). For $B\geq B_c$ the one-magnon cosine dispersion is shown.}
\label{transverse_contour}
\end{figure}

First, we note that the results for $S^{xx}(q,\omega)$ in Figs.
\ref{transverse_3d},\ref{transverse_contour} a) and b) are clearly
different from those for $S^{zz}(q,\omega)$ in Figs.
\ref{longitudinal_3d},\ref{longitudinal_contour} c) and d) at identical
magnetic fields. This is to be expected, since the application of a finite
magnetic field breaks the $SU(2)$ invariance of the AFHC. Second, long-wave
length transverse spin-excitations will experience the Zeeman energy due to
the magnetic field, which leads to a spin gap of size $B/J$ at $q=0$. This
has to be contrasted against the gap at $q=\pi$ in $S^{zz}(q,\omega)$ at finite
fields in Figs. \ref{longitudinal_3d}, \ref{longitudinal_contour} d).
Third, and as for the longitudinal case a field driven zero
mode at $q=q_s$ can be seen in \ref{longitudinal_contour} a), b) - with a
rather low intensity as $T\rightarrow 0$. In contrast to the longitudinal
case, this mode develops out of the zone zenter and moves to the zone
boundary with $q_s = 2\pi\langle S^z\rangle$
\cite{pytte_transversedispersion,ishimura_transversedispersion,Muller1981a}.

Even though it can be misleading to compare MaxEnt data based
on different QMC data sets quantitatively due to the underlying different
statistic quality, we notice enhanced spectral weight near the zone boundary
in Fig. \ref{transverse_3d} b) compared to zero magnetic field in Fig.
\ref{longitudinal_3d} b), which means that a weak uniform field strengthens the antiferromagnetic order in the transverse strucure factor. This effect was also observed in
\cite{Muller1981a} for small fields and by Karbach {\it et al.} within the
static structure factors \cite{karbachstatic}.

\subsubsection{The case $B\geq B_c$}

At intermediate fields selection rules \cite{Muller1981a} allow for a fairly
complex distribution of spectral weight as is
also obvious from the solid lines in Figs. \ref{transverse_contour} a) and
b). In contrast to this, above the saturation field and at low temperatures,
a straightforward picture emerges (see Fig.
\ref{transverse_3d},\ref{transverse_contour} d) and f)). In this regime and
for $T\rightarrow 0$ the systems is fully polarized. In that case the
elementary excitations are non-interacting ferromagnetic one-magnon states,
leading to a dispersion $E(k) = J \cos (k) + B$ in the transverse
structure factor, with a momentum-independent spectral weight
\cite{Muller1981a}. For finite $T$, we find that this picture is modified in
two ways. First, significant thermal broadening occurs, which as e.g. in
Fig. \ref{transverse_contour} e), at $B=2.5$ and $T=J$ can lead to a
complete closure of the zone boundary spin-gap. Second, and as can be seen
in Figs. \ref{transverse_3d} d) and f), there is a substantial wave-vector
dependence of the spectral weight in the cosine-signature of the one-magnon
state. The latter is due to the elementary one-magnon states being excitated
in a polarized background which contains thermal fluctuations
\cite{Grossjohann-unpublished}. Finally, we emphasize the difference
in the evolution of the overall spectral weight, contrasting longitudinal
versus transverse excitations. While in Fig. \ref{longitudinal_3d}
the weight of the excitations decreases with increasing
field, this is not so in Fig. \ref{transverse_3d}.

Figs. \ref{transverse_3d} and \ref{transverse_contour} bear a close resemblance to the concept of field-induced Bose-Einstein condensation of triplets, which has been under intense scrutiny for several quantum spin-systems recently \cite{Nikuni2000,Oosawa2001,Ruegg2003,Misguich2004,Sebastian2006,Chaboussant1997,Watson2001,Lorenz2008,Honda1998,Zapf2006,Zvyagin2007,Manaka1998,Garlea2007}.
These systems feature a gapful {\em zero-field} state with the
lowest triplet branch 'condensing' as the field is {\em increased}.
For the AFHC, this scenario is reversed, i.e. {\em decreasing} the field
through the critical value for complete polarization $B_c$, the magnons condense at $q=\pi$ and the system switches from a gapful state
to a Luttinger liquid of deconfined spinons. Obviously, the latter does not represent a true gauge-symmetry broken state, since (i) 1D-correlation functions decay algebraically and (ii) the magnons above $B_c$ are constrainted by a hard-core repulsion \cite{Affleck1990_91,Sorensen1993,Giamarchi1999}.

As the temperature is lowered, the thermal smearing of the approximately quadratic dispersion at $q=\pi$ for $B=B_c$ is reduced, see Figs.
\ref{transverse_contour} c) to d). For the momentum-integrated structure
factor this will lead to a critical increase of the density of states at
$\omega = 0$ as temperature is lowered. Clear indications of this critical
enhancement as $T\rightarrow 0$ have been observed in recent measurements of the transverse relaxation rate $T^{-1}_1$ in nuclear
magnetic relaxation experiments at $B=B_c$ on copper pyrazine dinitrate
Cu\-(C$_4$\-H$_4$\-N$_2$)\-(N\-O$_3$)$_2$ (CuPzN) \cite{Kuhne2008a}.

\section{Sum Rules}\label{D}

\begin{figure}[!ht]
\begin{center}
\includegraphics{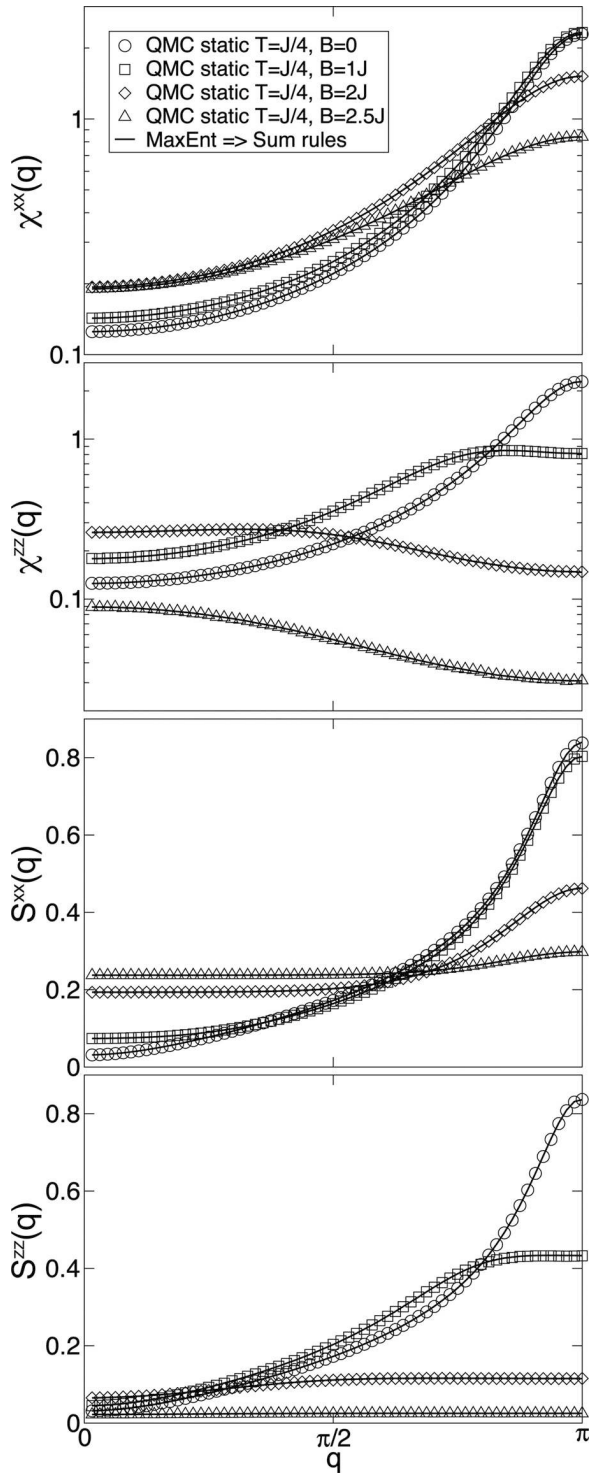}
\end{center}
\caption[1]{Comparison of transverse and longitudinal static susceptibility/structure factor (symbols) and sum rules (lines) for T=J/4 and four different magnetic fields $B/J=\left\lbrace 0, 1, 2, 2.5\right\rbrace $ (from top to bottom). All sum rule results are within the error bars of the static quantities which are within symbol size.}
\label{sumrules_all}
\end{figure}

Sum rules have been used extensively for the AFHC to evaluate the contribution
of 2- and 4-spinon excitations to the spectral weight of the dynamical structure factor at T=0 \cite{Karbach1997a,lakhalabada04,lakhalabada05,Caux2006a}.
For the present work sum rules can be applied to assess the quality
of the analytic continuation. We will focus on the sum rules for the static structure factor $S^{\alpha\beta}(q)$ and the static susceptibility $\chi^{\alpha\beta} (q)$ which obtained by integral transforming the dynamical structure factor \cite{hohenberg74}
\begin{eqnarray}
\label{eq11}
&& S^{\alpha\beta}(q) =  \frac{1}{\pi}\int_{0}^{\infty}d\omega(1+e^{-i\omega t})S^{\alpha\beta}(q, \omega)
\phantom{AAAAA}\\
\label{eq12}
&& \chi^{\alpha\beta}(q) = \frac{2}{\pi}\int_{0}^{\infty}d\omega \omega^{-1}(1-e^{-\beta\omega})S^{\alpha\beta}(q, \omega)\mathrm{.}
\end{eqnarray}
While $S^{\alpha\beta}(q, \omega)$ on the right-hand side of eqns.
(\ref{eq11}) and (\ref{eq12}) involve MaxEnt-data, the static structure factor
$S^{\alpha\beta}(q)$ in eqn. (\ref{eq11})
is calculated from a real-space Fourier transformation of the equal-time correlation functions and the static susceptibility $\chi^{\alpha\beta} (q)$ in eqn. (\ref{eq12}) can be evaluated from the Kubo integral
\begin{equation}
\chi^{\alpha\beta} (q) = \sum_r e^{iqr}\int_{0}^{\beta}d\tau \left\langle S^{\alpha}_r(\tau)S^{\beta}_0(0)\right\rangle \mathrm{.}
\label{kubo}
\end{equation}
of the imaginary time QMC-data. I.e. both, $S^{\alpha\beta}(q)$ and $\chi^{\alpha\beta} (q)$ are obtained from QMC-data which is {\em independent} from the MaxEnt continuation.
In particular the static susceptibility should provide for a clear consistency check regarding the low energy features in the zero field dynamic structure factor at $k=\pi$ as shown in Fig. \ref{k64}.

In Fig. \ref{sumrules_all} we compare the left- and right-hand sides of eqns.
(\ref{eq11}) and (\ref{eq12}) both, for the longitudinal and transverse components,
i.e. $\alpha\beta=zz$ and $\alpha\beta=xx$. First, we emphasize that the numerical values for $S^{zz}(0)$, $S^{zz}(\pi)$, $\chi^{zz}(0)$ and $\chi^{zz}(\pi)$ which we
have obtained at zero magnetic field
are consistent with those reported in refs. \cite{Starykh1997a, kimwiese} and
corroborate the parameters of scaling relations \cite{Starykh1997a}
$$S^{zz}(\pi) = D_s\ln(T_s/T)^{\frac{3}{2}},\quad D_s=0.094(1),\quad T_S=18.3(5)$$ 
$$\chi^{zz}(\pi)=\frac{D_{\chi}}{T}\ln(T_{\chi}/T)^{\frac{1}{2}},\quad D_{\chi}=0.32(1),\quad T_{\chi}=5.9(2)$$ for $T=J/4$. Second, Fig. \ref{sumrules_all} proves an excellent agreement of QMC data involving
analytic continuation to that free of the MaxEnt procedure. We have found
this agreement for all temperatures and all fields investigated, including
those not depicted here. All differences lie within the error bars of the static quantities which is remarkable, given that the typical MaxEnt error is
estimated to be $\sim$10-20\% \cite{Starykh1997a}. We note that we have performed this sum-rule check for various MaxEnt procedure, i.e. historic, classic, and bryan (see Appendix \ref{app1}) and found the same level of agreement.

\section{CONCLUSION}
\label{E}
In conclusion, using MaxEnt continuation of QMC results, we have analyzed
the evolution of transverse and longitudinal spin excitations of a AFHC
with 128 sites at finite temperatures and magnetic fields up to and above
the saturation field. Our results are consistent with and complement similar studies
using small system ED and zero-temperature BA. In particular we
have detailed the difference between longitutinal and transverse excitation
as a function of the magnetic field and temperature. Moreover we have
considered the field induced magnon 'condensation' at the saturation field
and the occurence incommensurate zero-modes. These investigation may be
of relevance to high-field NMR data on AFHC materials \cite{Kuhne2008a} as well
as to inelastic neutron scattering experiments.

Several open questions remain. While the issue of spin-diffusion has been
out of reach in this work, future analysis should improve the resolution
of the MaxEnt, in order to access the line-shapes at small $q$. This
also pertains to the form of the low-$\omega$ spectrum
of the zero-field dynamic structure factor at $q=\pi/2$. Finally
it will be interesting to perform similar calculations for various
generalizations of the AFHC including anisotropy and disorder.

{\it Acknowlegement} Part of this research has been funded by the DFG
through through Grant No. BR 1084/4-1. W.B. acknowledges the hospitality of the KITP, where this research was supported in part by the NSF under Grant No. PHY05-51164.

\appendix

\section{Maximum Entropy Method}
\label{app1}

In this appendix, and for completeness sake, we give a brief account of our
MaxEnt approach. In MaxEnt we minimize the functional \cite{gubernatis}
\begin{equation}
Q=\frac{1}{2}\chi^2 - \alpha S \,\,\,.
\label{maxent}
\end{equation}
For perfectly uncorrelated QMC data $\chi^2$ is the
least-square difference between the data $S^{\alpha\beta}(\tau)$, with standard deviation $\sigma_{\tau}$, and the
transform of the trial spectrum $A(\omega)$ to imaginary times using
the Kernel $K(\omega, \tau)$
\begin{equation}
\chi^2 = \sum_{\tau}\left[ \frac{S^{\alpha\beta}(\tau) - \frac{1}{\pi}\int_{0}^{\infty}d\omega K(\omega, \tau)A(\omega)}{\sigma_{\tau}}\right]^2
\label{chisq}
\end{equation}
where for brevity we disregard the q-dependency. In principle imaginary-time output from the QMC is correlated and needs to be transformed into an eigenbasis of the covariance matrix prior to using eqn. (\ref{chisq}) in order to work with decorrelated data. However, we have observed that diagonalizing the covariance matrix has neglegible impact on the spectra which we have analyzed. Therefore we have decided to ignore off-diagonal elements of the covariance matrix.

The second term on the right-hand side of eqn. (\ref{chisq}) contains the 
Shannon entropy 
\begin{equation}
S = \sum_{\omega}\left[ A(\omega) - m(\omega) - A(\omega)\ln\left(\frac{A(\omega)}{m(\omega)} \right) \right] \,\,\,,
\end{equation}
with respect to a default model $m(\omega)$ which prevents overfitting of
the data. We have used a simple flat default model for all calculations which was
iteratively adjusted to match the 0th moment of the trial spectrum. This is
different from ref. \cite{Deisz1990a}, where several sum-rules have been used
to apply additional bias to $A(\omega)$.

The choice of the Lagrange parameter $\alpha$ has been discussed extensively
\cite{gubernatis}. So-called {\em classical} and {\em historic} approaches use Bayesian logic to fix {\it one} $\alpha$ for the most probable spectrum $A_{\alpha}(\omega)$. More generally however, a probability distribution
$P\left[ \alpha | S^{\alpha\beta}(\tau) \right]$ exists \cite{gubernatis}, which
determines the most likely spectrum through the {\em average}
\begin{equation}
S^{\alpha\beta}(\omega) = \int d\alpha A_{\alpha}(\omega)P\left[ \alpha | S^{\alpha\beta}(\tau)\right]\,\,\,.
\end{equation}
We have analyzed our results in term of all three ways to choose $\alpha$. We found
classic and averaged spectra to be identical, indicating that $P\left[ \alpha | S^{\alpha\beta}(\tau) \right]$ is very sharp in our case, which supports the statistical quality of the underlying QMC data. As to be expected, for the
historic approach we found somewhat smoother results with a tendency to under-fit the data, thus all shown results are based on averaged spectra.

The minimization of (\ref{maxent}) is done via multi-dimensional Newtown iterations.
However, following Bryan's work \cite{bryan} we have reduced the effective search
directions by a singular value decomposition of the kernel
$K = U\Sigma V^{T}$
down to typically 10-20 of  the largest eigenvalues of $\Sigma$, depending on the temperature. This leads to a significant speed up of the algorithm.

\end{document}